# A Genetic Algorithm Approach to the Grooming of Dynamic Traffic in Tree and Star Networks with Bifurcation

Kun-hong Liu · Yong Xu · De-Shuang Huang

**Abstract** Traffic grooming is widely employed to reduce the number of ADM's and wavelengths. We consider the problem of grooming of dynamic traffic in WDM tree and star networks in this paper. To achieve better results, we used the bifurcation techniques to the grooming of arbitrary dynamic traffic in a strictly non-blocking manner in networks. Three splitting methods, including Traffic-Cutting, Traffic-Dividing and Synthesized-Splitting were proposed. A genetic algorithm (GA) approach based on these methods was proposed to tackle such grooming problems in tree and star networks. The performance of these algorithms was tested under different conditions in star and tree networks. Computer simulation results showed that our algorithm is efficient in reducing both the numbers of ADM's and wavelengths.

## 1. Introduction

Traffic grooming (TG) technique is an efficient way to reduce the total cost in today's backbone Wavelength Division Multiplexing (WDM) optical networks. In such networks, nodes are typically equipped with add/drop multiplexers (ADMs) to electronically combine low rate traffic streams onto a high rate wavelength to make the best use of wavelength's capacity. And how to intelligently assign traffic streams onto wavelengths to reduce the number of wavelengths and the usage of electronic equipments, such as ADMs, is what TG technique is going to resolve, which is unfortunately proved to be NP-complete [1] hence heuristics must be used.

Most of the previous work on TG was based on the ring and mesh topologies [1-11]. The TG problem in ring networks was first discussed due to their simplicity and widespread use in today's infrastructural networks [1-7]. Because mesh topology is more general and practical,

---

Kun-hong Liu[1,2] · Yong Xu[3] · De-Shuang Huang[1]

[1]Intelligent Computing Lab, Hefei Institute of Intelligent Machines, Chinese Academy of Sciences, P.O. Box 1130, Hefei Anhui 230031, China.

[2]Department of Automation, University of Science and Technology of China, Hefei, Anhui, 230026, China.

[3]School of Computer Science, The University of Birmingham, Edgbaston, Birmingham, B15 2TT, United Kingdom

e-mail:lkhqz@163.com

Y. Xu
e-mail:y.xu@cs.bham.ac.uk

D.S. Huang
e-mail:dshuang@iim.ac.cn

and the IP/WDM networks are mainly constructed based on such a topology, the grooming of traffic in mesh networks is becoming a prominent research area [8-11]. But it is obvious that other irregular topologies should also not be ignored, and each is of great value in their own sense. For example, star networks are widely deployed in LANs or MANs with a wide area backbone, whereas most cable TV networks and PONs are based on tree topologies. But there are very few papers discussing TG problems in stars and trees [3, 12-15].

The static TG problem in star network is NP-complete，as proved in [13], so the problem of grooming arbitrary dynamic traffics in star topology is much more difficult, no to mention the tree topology. Hence it is no wonder that the previous work mainly considered the static TG problem in star networks [3, 12-14]. Because traffic often changes frequently, grooming of arbitrary dynamic traffic will be of more practical value. The grooming of dynamic traffic can be viewed as the grooming of a set of traffic patterns, which was first addressed by Berry and

Modiano in [4]. There are mainly two classes of the grooming of dynamic traffic [5]: *strictly nonblocking* and *rearrangeably nonblocking*. In this paper, we focus on the strictly nonblocking grooming problem, because in this way, all traffic demands in a new traffic pattern can be established without being interrupted, which is realized by assigning the same traffic demand in different traffic patterns to the same wavelength. The rearrangeably nonblocking grooming, which means that each traffic demand can be assigned to the different existing wavelengths with the set of nodes each wavelength drops at kept unchanged for each new traffic pattern, may result in saving more ADMs compared to those of the strictly nonblocking grooming [7], and it will be our future work.

To the best of our knowledge, we are the first to analyze the problem of strictly nonblocking grooming of arbitrary dynamic traffics in star and tree networks without bifurcation in [15]. Although bifurcation is an efficient way for further saving of ADMs and wavelengths, it is not easy to work out a good solution by splitting traffic flows because it will inevitably make the problem much more complex. We proposed three splitting methods in [6, 7]: Traffic-Cutting, Traffic-Dividing and Synthesized-Splitting. These methods can surely result in further reducing the number of ADM's and wavelengths for strictly nonblocking and rearrangeably nonblocking grooming in rings respectively. In this paper, we will apply these splitting methods to resolve the TG problem in tree and star networks, and the approach is also based on GA.

Because the TG problem in trees are NP-complete even when every interior node has full wavelength conversion capability [13], the authors of this paper first decomposed a tree topology into stars, then proposed a greedy heuristic for grooming of static traffic in star networks to tackle the TG problem in tree networks. It is obvious that this two-stage solution would inevitably be complex and very time-consuming. What is more, as the greedy heuristic would depend on the order of traffic streams being routed, it is hard to get globe optimism for all cases even if the solution in each star can reach its optimal. Therefore, we propose a GA approach which can groom dynamic traffic in tree topology with bifurcation in one step, and the algorithm is out of the influence of order of traffic streams. Furthermore, the new algorithm can split traffic streams and then groom them in trees so as to further save ADMs and wavelengths. As a star network can be considered as a tree with only one interior node, the algorithm can be applied to star networks with slightly adaptation. Our main objective is to minimize the total number of ADM's, and the secondary goal is minimize the total number of wavelengths, and they are quite different from the goal in [13] which is to minimize the total amount of electronic switching.

The rest of this paper is organized as follows. We give the general idea of TG problem in trees and stars respectively, and propose the three splitting methods in Section 2. In Section 3, we describe the GA approach, and Section 4 gives the computer simulation results along with corresponding discussions. In the last Section, we conclude this paper.

## 2. Problem Definition and the Splitting Methods

We propose a genetic algorithm to deal with the TG problem in star and tree networks. The networks considered in this paper consist of $n$ nodes numbered by 0, 1, …, $n$-1, and each node is equipped with a number of ADMs to add/drop traffic demands at it. We present the dynamic traffic demands by a set of $n \times n$ traffic matrix $\boldsymbol{R}=\{\boldsymbol{R}^m\}$ ($m$=0, 1, …, $M$-1), each of which represents the traffic demands at time $t=m$. Each element $r_{ij}^m$ of the traffic pattern $\boldsymbol{R}^m$ represents the traffic demand originating from node $i$ and terminating at node $j$ at the moment $t=m$, and the traffic patterns in this set are activated one by one with time changing. In [6, 7], we proved that by splitting traffic demands in rings, the ADMs and wavelengths can be further saved. In this paper, we apply this technique in star and tree networks to tackle the TG problem with bifurcation in a strictly nonblocking manner. The following problems are discussed under the assumption that the maximum traffic demands are not larger than the capacity of a wavelength.

2.1. Problem Definition

We discuss the TG problem based on the network model we proposed in [15], and for the integrity of the discussion, we give a brief description here.

Tree networks can be classified into to categories: regular tree and irregular tree. The main difference among the tree networks lies in the construction as there is only one route between each node pair. This is significantly different from bidirectional ring and mesh networks and therefore, and we can safely ignore the routing problem. If a traffic demand $r_{ij}$ is going to be assign to a wavelength, we only need to take the following two steps:
1. find the route between node $i$ and $j$;
2. judge whether all the links among the route have spare capacity to accommodate this traffic.

After the relationship between each pair of nodes on the tree is defined, a route between different node pairs can be established by tracing a node's father or child node. Then by setting up a minimum subtree which contains two end nodes, a lightpath can be established. Since the process of establishing a route is similar in all kinds of tree networks, it is clear that the way of constructing of a tree topology will not affect the performance of our algorithm, and the algorithm can be applied to all kinds of tree networks with slight adjustment in the description of the tree topology. For simplification, we present our algorithm in binary tree networks.

In our tree networks, all the links are bidirectional. Each internal node is designed as a SONET-over-WDM model, as proposed in [8] and equipped with SONET components (DCS, OXC, etc.), which can provide fast multiplexing/demultiplexing capability. Each leaf node is equipped with a wavelength add/drop multiplexer. In this way, we will only add an ADM for each wavelength at each internal node and leaf node if the traffic needs to be added/dropped at the node, and need not consider other equipments. The root node of the tree is labeled as node 0. And we regard a star network as a tree with only the hub node as its internal node, which also labeled as 0 node. It is not necessary to discuss the routing problem when assigning traffic to a wavelength in tree topologies. If a traffic demand $r_{ij}$ is going to be assign to a wavelength in tree networks, we may search the route between a node pair node $i$ and $j$, then judge whether all the links among the route have spare capacity to accommodate the traffic. For star networks, we need the second step. In this way, the algorithm designed for tree networks can be applied to star networks with only slight adjustment.

## 2.2. Splitting Methods

Bifurcation means to divide a traffic stream into smaller parts or shorter segments so as to fill them into the links with spare capacities in a wavelength. To realize bifurcation in WDM ring networks, we proposed three splitting methods in [6]: Traffic-Cutting, Traffic-Dividing and Synthesized-Splitting. As was proved in [6, 7], we can get better results by applying these three splitting methods in both strictly nonblocking and rearrangeably nonblocking grooming of traffic. Obviously, these methods are not constrained to ring topology. In this paper, we apply them to solve the TG problem in a strictly nonblocking way in star and tree networks. It should be noticed that in strictly nonblocking grooming, the same divided parts or cut segments of a traffic demand at different moment must be assigned to the same wavelength.

### 2.2.1. Traffic-Cutting

In [6, 7], the Traffic-Cutting method is to cut a traffic flow into short segments and assign them to fit the "gaps" of existing wavelengths separately. In this way, the number of ADM's can be reduced by sharing these cut segments with the ADMs in the existing lightpaths, and the number of wavelengths can also be cut down as a byproduct with more efficient utilization of the spare capacities in each link.

The conditions needed for cutting a traffic flow is different in [6] or [7]. In this paper, a traffic flow is cut only when the following two conditions are both satisfied:

*Condition 1*. One of the end nodes of the traffic flow is already the dropping node of the current wavelength, and the wavelength has another dropping node $f$ between the two end nodes of the traffic; at the same time, node $f$ and the other end node are already dropping nodes of an already existing wavelength.

*Condition 2*. The two cut segments can be assigned to the two wavelengths respectively.

These two conditions can make sure that when a traffic flow is cut, the two cut segments can be assigned to the current wavelength and an existing wavelength without additional ADMs. By sharing the existing ADMs with other traffic demands on these two wavelengths, we can assign the two cut segments to the spare link loads. In order to ensure that a traffic flow be cut more efficiently, we always try to cut a traffic flow to fit into the current wavelength at the largest possible length. Then the remaining segment will be as short as possible, so it can be fit onto the existing wavelengths more easily.

Since cutting a traffic flow will make it pass through two or more wavelengths and become multi-hopped though the original connection is only single-hopped, the control of a network will become more complex and the signal's transmission will be inevitably delayed when it passes on different wavelengths. So in this paper, a traffic flow will be cut into two segments at most so as to make the process acceptable.

It is obvious that this method can not be applied to the traffic demands which pass only one physical link. So there are $n$-1 traffic demands which can not be cut.

### 2.2.2. Traffic-Dividing

As was proved in [6, 7], it is an efficient way to save ADMs by dividing a higher-rate traffic flow into a few lower-rate parts, and assigning these parts into the spare capacity of existing wavelengths. In this process, the traffic flow is assigned to different wavelengths by sharing the ADMs with the existing traffic demands. Similar to the Traffic-Cutting method, this method will be applied only when the following two conditions are satisfied at the same time:

*Condition 3*. Both of the end nodes of the traffic flow are already the dropping nodes of the current wavelength.

*Condition 4*. The one of the divided parts can be assigned to that wavelength.

These conditions are used to guarantee that one divided part of the traffic can be assigned to the current wavelength without adding a new ADM, then the number of ADM's and wavelengths can be reduced simultaneously. Only when none of the links' load that the traffic may pass through is full, can we divide it. So we must check the current wavelength to ensure that the divided part can be accommodated by the wavelength. The way of dividing a traffic is decided by the remaining capacity of a wavelength: the divided part which will be assigned to the current wavelength is equal to the minimum spare capacity of its passing route on the wavelength, and another remains in the traffic matrix and will be assigned to other wavelengths. Since the rate of the remained part is lower than the original traffic, it has higher chance to be assigned to the existing wavelengths to which the original one can not be assigned.

Different from the first splitting method, this method will assure that each part of the divided traffic is still single-hopped and the traffic transmission will not be delayed although a traffic flow may pass through two or more wavelengths after divided. And this method can be applied to all of the traffics, including the physical single-hopped traffic demands. But a traffic flow can not be divided into too many parts either, otherwise it would be more difficult to manage the network.

2.2.2. Synthesized-Splitting

As the above two splitting methods are both efficient in saving ADM's and wavelengths, we design the third splitting method by combining them together, which is named as the Synthesized-Splitting. When this method is applied, we divide a traffic flow first, then try to cut it if possible, so the traffic becomes shorter in length as well as lower in rate. Then the traffic will be much easier to be fit into the wavelengths, and this method can lead to best results in most cases as was demonstrated in [6, 7]. We will apply this method to star and tree networks in this paper. When utilizing this method, the traffic flow can be divided into two or more parts first. After divided, if the remaining parts in the traffic matrix can be cut, we only cut a traffic into at most two segments, and the two cut segments are assigned to two wavelengths at the same time and will no longer be divided. All these operations are done only when the corresponding conditions are satisfied.

## 3. Genetic Algorithm for Solving Grooming Problems in Tree Networks

In this paper, we will propose a GA approach to tackle the TG problems in star and tree networks. The framework of this GA approach is based on that described in [15]. For star and tree networks, the GA framework is the same except for decoding approaches. With the three splitting methods described in Section 2 imbedded in this algorithm, the local improvement algorithms are much more efficient in saving ADMs and wavelengths. We use the $(\mu+\lambda)$-strategy to produce offspring, as described below:

1. Set all necessary parameters, and let $t=0$;
2. Generate an initial population $P_0$ with $\mu$ different chromosomes at random;
3. REPEAT:
    a. Apply crossover and mutation to the parents to produce $\lambda$ offspring;
    b. Decode each chromosome $i$ ($i=1, 2, \ldots, \mu+\lambda$) with the splitting methods to assign each traffic demand in the $M$ traffic patterns into wavelengths in a strictly nonblocking way;
    c. Evaluate each individual $i$ ($i=1, 2, \ldots, \mu+\lambda$) in both the parents and the offspring, and select $\mu$ individuals with the highest fitness value for the next generation;
    d. Set $t=t+1$;
4. Until some "termination criterion" is satisfied.

The realization of each step is described in the following subsections.

### 3.1. Chromosome Representation

The $M$ traffic matrices representing the traffic request in different time slots are converted into $M$ $n(n\text{-}1)$-dimensional vectors, $X^m = (x_1^m, \ldots, x_k^m, \ldots, x_{n(n-1)}^m)$, $m=1, 2, \ldots, M$. A random permutation of $N= n(n\text{-}1)$ different integers is generated in the range $[1, N]$ to represent a random permutation of the traffic elements, and will be decoded with the $M$ traffic matrices in a strictly nonblocking grooming manner described in Algorithm I.

### 3.2. Decoding Approaches and Fitness Assignment

Because the same traffic element in different traffic patterns must be assigned to the same wavelength in strictly nonblocking grooming, we decode the chromosome with each of the $M$ traffic patterns one by one. In order to minimize the number of ADM's and wavelengths, we try to assign the traffic with the same end nodes into the same wavelength. In this way, if a wavelength drops at $\alpha$ nodes, it can accommodate at most $\alpha(\alpha\text{-}1)$ traffic demands. Based on this fact, we propose a first-fit approach incorporated with a greedy improvement to decode chromosomes. In this approach, a traffic item $x_k$ is assigned to a wavelength only if all the traffic demands represented by it in the set of $M$ traffic patterns can be assigned to it. In this way, the strictly nonblocking grooming can be realized.

The process of decoding works as follows: Firstly, the first encountered traffic item $x_i$ in the chromosome is assigned to a wavelength. Then, it examines the remaining traffic items one by one to see whether they can be assigned to the existing wavelengths with an additional ADM. When such a traffic item $x_j$ is found, it is assigned to the existing wavelengths, and the two items ($x_i$ and $x_j$) are exchanged. When all the items are checked and none can be assigned to the current wavelength, this step stops. Thirdly, the algorithm checks the traffic item $x_k$ in the chromosome again. If a traffic $x_l$ can be assigned to the current wavelength without adding additional ADMs, assign it to the current wavelength then exchange the two traffic items ($x_k$ and $x_l$) are. This process continues until all the traffic items in the chromosome are examined. Fourthly, when no traffic can be assigned to the current wavelength as a whole, the local improvement algorithm is called to examine the traffic items one by one to find whether a traffic item can be assigned to it by he splitting methods without adding ADMs. This step stops when all the traffic items have been checked. Then the algorithm proceeds with the next wavelength and repeat the above three steps, until all the traffic items have been assigned to appropriate wavelengths.

We used the so-called wavelength reuse technique proposed in [15] in this algorithm, which means to check whether the existing wavelengths can accommodate a traffic item before it is assigned to the current wavelength. This is necessary in irregular topologies for better utilization of the spare capacity of a wavelength, which can lead to further saving in ADM's and wavelengths in most cases. Splitting methods are also used in this algorithm to accomplish more traffic items to the existing wavelengths to achieve further better results, with which the performance of the local improvement algorithm will be better than that in [15].

Algorithm 1 implements the decoding process. It can serve as a solution for the traffic assignment in strictly nonblocking grooming way in star and tree networks. When applying this algorithm to optimize the traffic assignment in different irregular networks, the way of traffic assignment need be adapted according to the network structure.

**Algorithm 1. Decoding A Chromosome**

Step 1. $k=0$ and $w=1$;

Step 2. If each of the $M$ traffic demands represented by traffic item $x_k$ can be assigned to wavelength $w$, assign each of them to it. Otherwise, go to Step 9;

Step 3. $f=1$, $l=k+1$;

Step 4. If each of the $M$ traffic demands represented by traffic item $x_l$ can be assigned to wavelength $f$ with only one additional ADM added to it, assign each of them to it, $l=l+1$, exchange $x_k$ with $x_l$, $k=k+1$;

Step 5. $f=f+1$, If $f<w$, go to step 4;

Step 6. If each of the $M$ traffic demands represented by traffic item $x_l$ can be assigned to wavelength $w$ without additional ADM added to it, assign each of them to it, $l=l+1$, exchange $x_k$ with $x_l$, $k=k+1$;

Step 7. $l=l+1$. If $l< n*(n-1)$, go to Step 6;

Step 8. $i=k$;

Step 9. If each of the $M$ traffic demands represented by traffic item $x_i$ can be divided into two lower-rate parts according to Conditions 3 and 4 given in Section 2, divide each of the traffic demand and assign a part of them to $w$; the other part is left in the traffic matrix;

Step 10. $i=i+1$. If $i< n*(n-1)$, go to Step 9;

Step 11. $j=k$;

Step 12. $s=1$;

Step 13. If there is a dropping node between the two end nodes, and each of the $M$ traffic demands represented by traffic item $x_j$ can be cut into two shorter segments and assigned to current wavelength and another wavelength $s$ according to Conditions 1 and 2 given in Section 2, cut each of the traffic demand and assign the two segments to $w$ and $s$, set $x_j=0$, exchange $x_k$ with $x_j$, $k=k+1$;

Step 14. $s=s+1$. If $s<w$, go to Step 13;

Step 15. $j=j+1$. If $j< n*(n-1)$, go to Step 12;

Step 16. $k=k+1$. If $k< n*(n-1)$, go to Step 2, otherwise, stop;

Step 17. $w=w+1$, go to Step 2;

In this algorithm, steps 1-7 and 16-17 constitute a grooming algorithm without bifurcation; steps 8-10 are employed to accomplish the Traffic-Dividing method and steps 11-15 are employed to accomplish the Traffic-Cutting method. If steps 8-15 are executed in sequence, the Synthesized-Splitting method is accomplished.

In detail, step 2 tries to assign the first-encountered traffic to current wavelength; step 4 reuse wavelength by trying to assign the followed traffic items to the existing wavelength adding at moat one ADM; step 6 examines the remaining traffic items one by one and tries to assign proper ones whose end nodes are already the dropping nodes on the current wavelength, so as to minimize the number of ADM's; step 9 divides traffic flows and assigns a part of them to the current wavelength without adding ADMs; step 13 is to cut traffic flows and assign them to two wavelengths with no additional ADMs; step 17 starts a new wavelength. This decoding process applies to all the chromosomes, and after assigning the $M$ traffic patterns to wavelengths, the numbers of ADMs and wavelengths can be calculated, then an individual's fitness value can be calculated accordingly.

For each individual, the fitness value is determined by the number of ADM's first. The fewer ADMs one requires, the higher fitness value it gets. If two individuals require the same number of ADMs, the fitness value will be decided by the required number of wavelengths, and the one requiring fewer wavelengths gets higher fitness value. If the number of wavelengths is still the same, a same fitness value will be assigned to them. For the all individuals, only the $\mu$ individuals with the highest fitness value will survive for the next generation.

By keeping the individuals with the highest fitness value for the next generation, the whole population will finally be led to evolve toward the global optimum rapidly with the local improvement. We will show the effectiveness of this algorithm by computer simulations in Section 4.

### 3.3. Crossover and Mutation

In crossover, we used the Order-Mapped Crossover (OMX) operator proposed by Xu and Xu [16] which can guarantee that no illegal offspring will be produced, and the offspring is able to preserve the ordering message from their parents.

In mutation, we adopt the simple inversion mutation. This operator randomly selects two points in a parent then an offspring is produced by inversing the genes between the two points.

## 4. Computer Simulations and Discussions

The performance of the proposed strictly nonblocking grooming algorithm with three splitting methods was tested in star and tree networks respectively under different conditions. The traffic demands from node $i$ to node $j$ ($i, j = 0, 1, \ldots, n-1$) are random integers distributed uniformly in the interval [0, 15]. The numbers of traffic patterns are 2, 4 and 8 respectively. In all the simulations, we set the population and the offspring sizes to 200 and the algorithm stops after 500 generations. And we set the crossover and mutation rates to 0.6 and 0.4 for every individual, and the two operators operate in consequence. The initial populations are randomly generated. All the problems are tested 10 times, and only the best results are shown. For the traffic patterns, we first generate two random traffic patterns $\boldsymbol{R}^1$ and $\boldsymbol{R}^M$ to represent two extreme traffics, then all the other traffic patterns $\boldsymbol{R}^m$ ($1<m<M$) are generated with each element $r_{ij}^m$ the random integer between $r_{ij}^1$ and $r_{ij}^M$, i.e., $r_{ij}^m = \text{random}[r_{ij}^1, r_{ij}^M]$, representing the traffic demand originating from node $i$ and terminating at node $j$. When the number of nodes are the same, the two extreme traffic patterns $\boldsymbol{R}^1$ and $\boldsymbol{R}^M$ are the same in all the tests.

We use the bounds proposed in [15] to evaluate the performance of our algorithm. We assume that two nodes $i$ and $j$ are directly connected and form a father-child node pair, and $i$ is the father and $j$ the child. $L_{ij}^{1m}$ donates the link load in the links which carry traffics from node pair $i$ to $j$ for traffic pattern $m$ in direction 1, and $L_{ij}^{2m}$ donates the link load in the links carrying traffic from node $j$ to $i$ in direction 2. The maximum link load of all the traffic patterns in the two directions between node pair $i$ and $j$ are defined as $L^1_{ij}$=max$_{x=0,...,M-1}$($L_{ij}^{1x}$) and $L^2_{ij}$=max$_{x=0,...,M-1}$($L_{ij}^{2x}$) respectively. Let $\sigma^m(s)$ denote the total traffic for traffic pattern $m$ dropped at an internal node $s$, and $\tau^m(s)$ denote the total traffic added at this node, then by defining $\sigma(s)$=max$_{x=0,...,M-1}$($\sigma^x(s)$), $\tau(s)$=max$_{x=0,...,M-1}$($\tau^x(s)$), and the set of the internal nodes in the tree as $\lambda$, the minimum number of wavelengths is given by

$$W_{\min} = \max_{i,j=0,...,n-1; i \neq j, s \in \lambda} (\lceil L_{ij}^1/g \rceil, \lceil L_{ij}^2/g \rceil, \lceil \max(\sigma(s),\tau(s))/g \rceil) \quad (1)$$

The minimum number of ADM's is given by

$$M_{\min} = \sum_{i=0}^{n-1} \lceil Max(\sigma(i),\tau(i))/g \rceil \quad (2)$$

The maximum number of wavelengths on binary tree networks is given by

$$W_{\max} = \begin{cases} (n^2-1)/4 & n \text{ is odd} \\ n^2/4 & n \text{ is even} \end{cases} \quad (3)$$

The maximum number of wavelengths on star networks is given by

$$W_{\max} = n\text{-}1 \quad (4)$$

The maximum number of ADM's on binary tree networks is given by

$$M_{\max} = n(n\text{-}1) \quad (5)$$

The maximum number of ADM's on star networks is given by

$$M_{\max} = n \times W_{\min} \quad (6)$$

Fig. 1 shows the grooming results versus the number of nodes on trees with 2 and 8 traffic patterns when the traffic granularity $g$=16. We can see from Fig. 1 that the results without splitting are the worse in all the cases, and those of the Traffic-Dividing and the Synthesized-Splitting methods are much better usually. Unlike the observation in [6, 7], the Traffic-Cutting does not perform as good as the other two splitting methods because the conditions for cutting are different from those in [6, 7]. And with the increase of the number of traffic patterns, its results become much worse than the other two splitting methods. This is because when $g$ is small there are very few ADMs per wavelength in average. What is more, although the wavelength reuse technique is applied to search the existing wavelengths and try to assign the traffic items to them with adding one additional ADM, this technique add only a few ADMs to a wavelength. As the Traffic-Cutting method requires that after cutting a traffic flow, the two cut segments are assigned to the current wavelength and an existing wavelength at the same time without adding ADMs, the probability of satisfying the two conditions of this method is very low especially when there are a lot of traffic patterns. For example, when $g$=16, with 8 traffic patterns and 15 nodes in a tree, after proper arranging ADMs and wavelengths, there are only 3.7 ADMs per wavelength without splitting. At the same time, the maximum traffic demand is equal to granularity and the average traffic demands are larger than half of granularity. Then not too much spare link capacity is available. As a result, at most 19 in totally 210 traffic demands will be cut in a run. So with this method only one more ADM and no more wavelength can be saved compared with the results of non-splitting. In contrast, 51 traffic demands can be cut at most when there are two traffic patterns, which results in 6 ADMs and 5 wavelengths being saved at most.

The Traffic-Dividing method tries to assign a part of a traffic demand to the current wavelength, and the remaining part will be assigned to the existing wavelengths with the aid of the wavelength reuse technique. So this method can result in further saving in the number of ADMs and wavelengths even when the traffic changes frequently, and perform better than what was shown in [6, 7] because the wavelength reuse technique did not be applied in ring networks. We find that when there are 2, 4 and 8 traffic patterns, this method can lead to 3, 5, and 11 ADMs and 8, 16, and 8 wavelengths savings respectively, which is 1.77%, 3.17% and 5.29% and 19.0%, 25.5% and 14.3% in the saving of ADM's and wavelengths compared with the non-splitting method. So it can be concluded that its performance will not be influenced by the number of traffic patterns obviously. In fact, when traffic changes violently, the non-splitting method can not achieve good results, so the advantage of this method becomes evident.

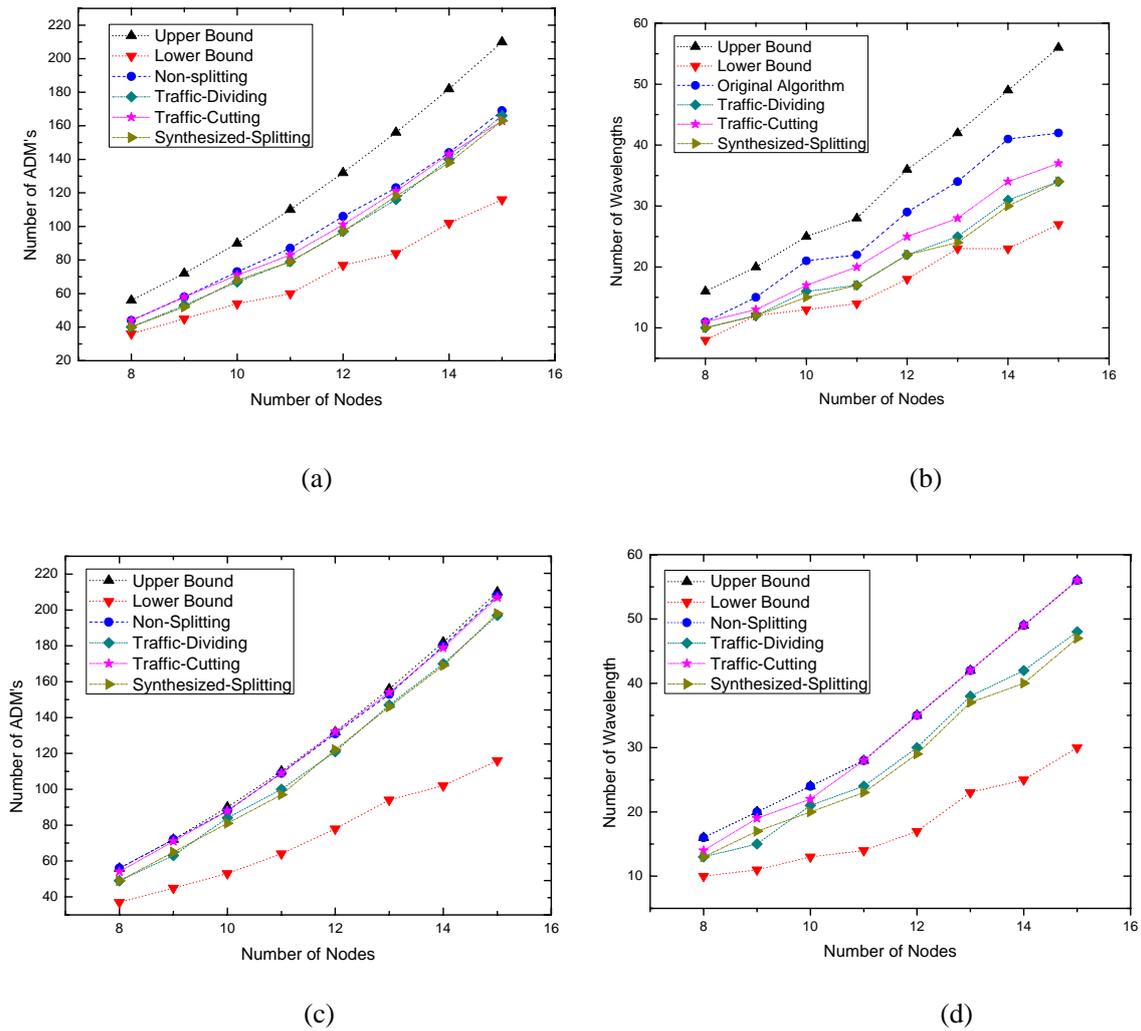

**Fig. 1** Computer simulation results for different number of nodes when g=16 in tree networks. (a) The number of ADMs vs. nodes for M=2; (b) The number of wavelengths vs. nodes for M=2; (c) The number of ADMs vs. nodes for M=8; (d) The number of wavelengths vs. nodes for M=8.

As was observed in [6, 7], the Synthesized-Splitting method achieves the best results in most cases at the cost of consuming much more time than other two splitting methods. Although it can not cut more traffic flow than the Traffic-Cutting method, by combining the two methods together, it can usually save some more, though not too many, ADMs and wavelengths than other methods. For example, when there are two traffic patterns, it can save 3 and 2 more ADMs and 0 and 1 more wavelengths compared with the results of the Traffic-Dividing method when 15, 14 nodes in a tree respectively. We can see from Fig.1, that the curves represented the results of this method are usually the closest to the lower bounds. The most obvious case is when there are 11 nodes and 8 traffic patterns, it can further save 12 ADMs and 5 wavelengths compared to the non-splitting method in a tree, which is 11.0% in the saving of ADM's and 17.8% in the saving of wavelengths'. Under the same condition, the Traffic-Cutting method can not achieve any saving, and the Traffic-Dividing method can achieve 9 ADMs and 4 wavelengths.

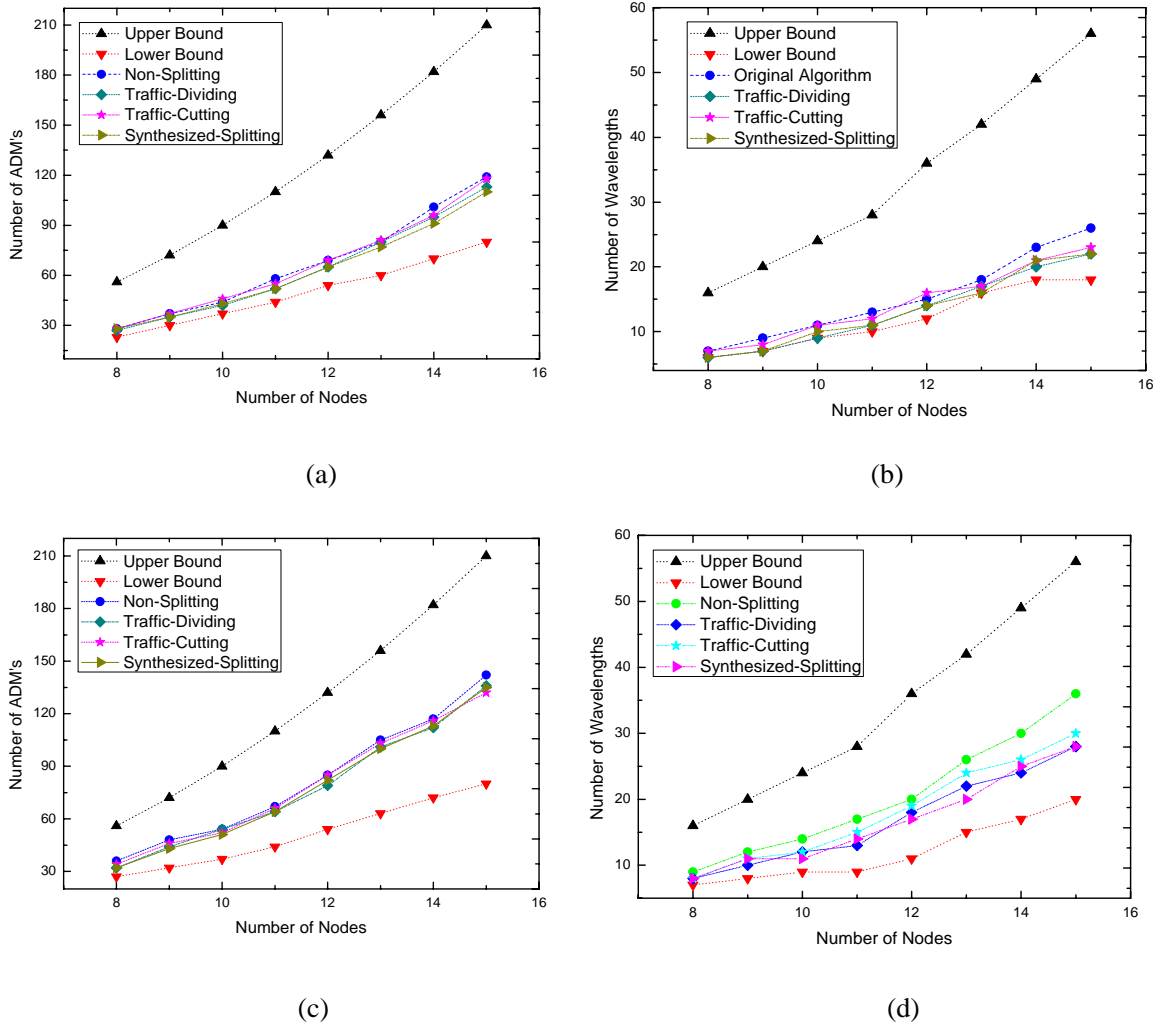

**Fig. 2** Computer simulation results for different number of nodes with g=24 in tree networks. (a) The number of ADMs vs. nodes for M=2; (b) The number of wavelengths vs. nodes for M=2; (c) The number of ADMs vs. nodes for M=8; (d) The number of wavelengths vs. nodes for M=8.

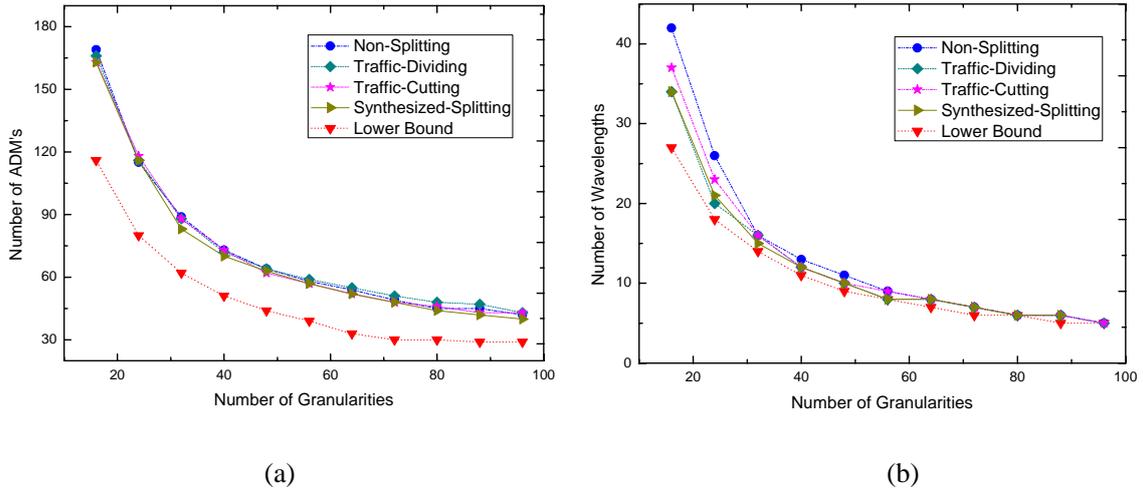

**Fig. 3** Computer simulation results for different granularity with 2 traffic patterns and 15 nodes in trees. (a) The number of ADMs vs. nodes; (b) The number of wavelengths vs. nodes.

The best results for different nodes and traffic patterns with g=24 on tree networks are shown in Fig.2. It can be found that when the traffic granularity is larger, there are more ADMs in a wavelength since more traffic demands can be assigned to it. We find that when there are 8 traffic patterns and 15 nodes in a tree, there are 3.9 ADMs per wavelength without splitting. At the same time, more spare links may as well be available in a wavelength because the granularity is larger than the maximum traffic demands. So it is easier to satisfy the two conditions of the cutting method, and at most 35 traffic demands will be cut, which is far more than 19 traffic demands when g=16. With the increase of granularity, the Traffic-Cutting method can perform better by cutting more traffic flows. When there are 2, 4, and 8 traffic patterns, with this method, 1, 5, and 10 ADMs and 3, 4, and 6 wavelengths can be further saved, which is a great improvement compared with those with g=16. With the same condition, the Traffic-Dividing method can save 6, 6, and 6 ADMs and 4, 6, and 8 wavelengths, and the Synthesized-Splitting method can save 9, 7, and 7 ADMs and 4, 8, and 8 wavelengths. So it seems that these two methods perform steadily regardless of the numbers of traffic patterns and can lead to better results than those with the Traffic-Cutting method in most cases. Since the non-splitting method is already a good algorithm, these three methods can not always get more saving because their search space is much larger than that of non-splitting. But in most cases, they can surely save more ADMs and wavelengths. In addition, it should be noticed that the saving achieved by these three methods will be more obvious when the number of nodes in a tree is large enough.

Although our main purpose is to save some more ADMs, we find that the saving of wavelengths is much more noticeable with each method in all the cases, and it may reach more than 20% saving in some cases. This is because by assigning more traffic parts or segments to a wavelength, the wavelength's capacity is much more fully occupied than without splitting, which results in the reduction of the wavelength usage. Therefore, compared to the ADM's, the saving in wavelength is much more evident.

Fig. 3 gives the computer simulation results for different granularity with 2 traffic patterns and 15 nodes in trees. It is obvious that with the increase of *g*, the number of ADM's and wavelengths decreases. The results of the Traffic-Cutting method come closer to those of the non-splitting method, and the other two methods get close results and are not far from the lower bounds. Generally, the results of these three splitting methods are better than those without splitting, and the Synthesized-Splitting method still achieves the best results. For example, though the number of wavelengths are all equal to the lower bound when g=96, it can still save 2 ADMs at most, while other methods can no longer save any ADMs.

Fig. 4 gives the simulation results for different number of nodes with 2 and 4 traffic patterns in star networks. Similar conclusions can be drawn from it. As is shown in Figs. 4(a) and 4(b), when the granularity is 16, the Traffic-Cutting method can not achieve any further saving in most cases, so most of its results and those without splitting are quite close or equal to the upper bounds. This is because of the violent changes of the traffic and the unbalance of the two directions of the fibers in the bidirectional star, which would unavoidably lead to consume more ADMs. Furthermore, in star networks, a traffic flow may arrive its destination only passing through an internal node, the hub node. So the Traffic-Cutting method can only cut a traffic flow at that node, which greatly limits its performance. We find that it can at most cut 12 traffic flows even when there are 15 nodes in a star. Moreover, we can see from these figures that the upper and lower bounds on the number of ADMs are so close to each other. As a result, it is very difficult to save more ADMs by this method. But it can result in saving one wavelength in some cases in spite of the fact that the required number of wavelengths is already small. What is more, the average results of this method are usually better than those of non-splitting.

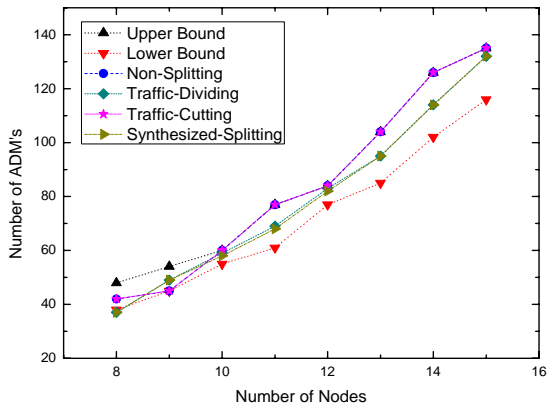

(a)

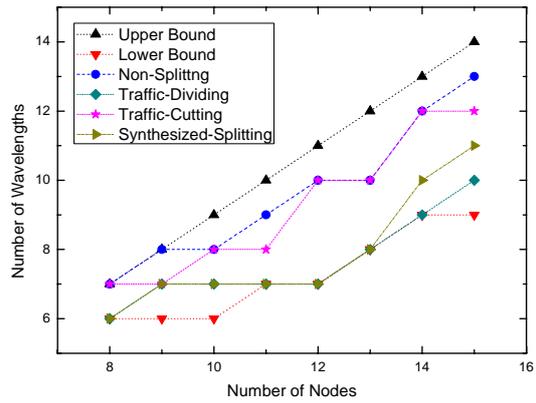

(b)

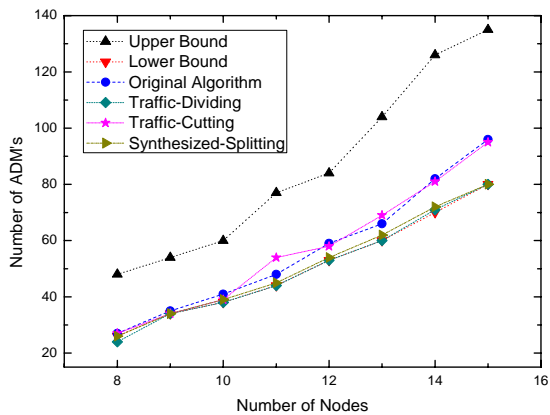

(c)

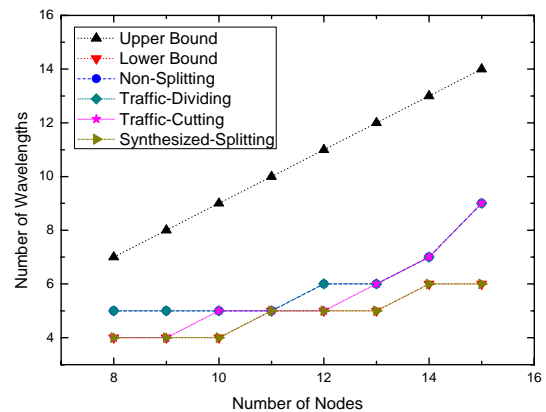

(d)

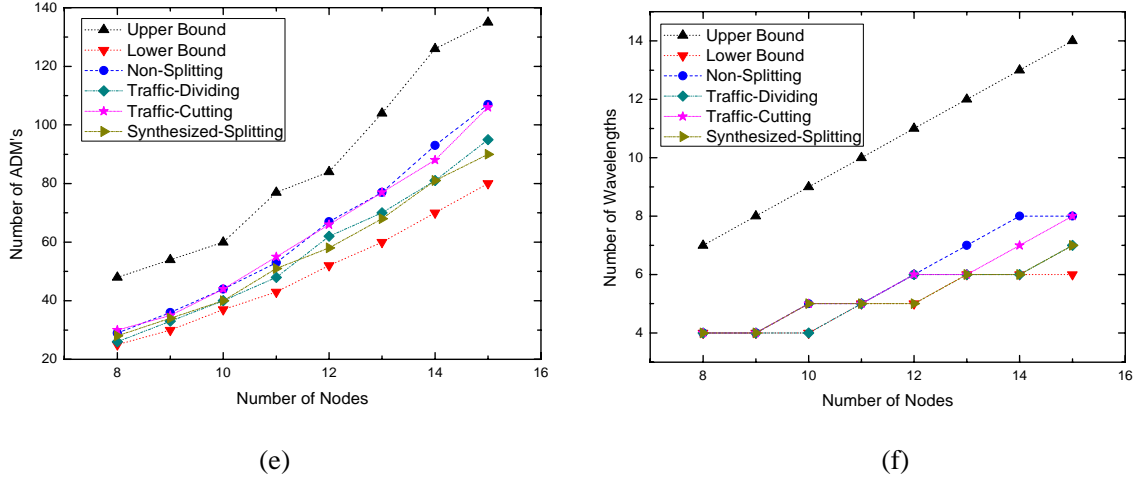

(e)                 (f)

**Fig. 4** Computer simulation results for different number of nodes with 2 or 4 traffic patterns in star networks. (a) The number of ADMs vs. nodes for g=16 and 2 traffic patterns; (b) The number of wavelengths vs. nodes for g=16 and 2 traffic patterns; (c) The number of ADMs vs. nodes for g=24 and 2 traffic patterns; (d) The number of wavelengths vs. nodes for g=24 and 2 traffic patterns; (e) The number of ADMs vs. nodes for g=24 and 4 traffic patterns; (f) The number of wavelengths vs. nodes for g=24 and 4 traffic patterns.

Fig. 4 shows that the Traffic-Dividing method and Synthesized-splitting method can lead to better results in nearly each case. When g=16, these two methods can at most save 3 and 3 ADMs and 3 and 2 wavelengths respectively. Our simulations showed that when there are 15 nodes in a star network, the average link load is 13.6 for the first traffic pattern, 12.3 for the second with only 10 ADMs in each wavelength in average. With such a small granularity, there are not too much spare capacity to be utilized for the splitting methods, so it is understandable that the three splitting methods can not achieve better saving.

When $g$ is equal to 24, from Figs. 4(c) to 4(f), it is clear that the number of ADM's and wavelengths are all very close the lower bounds, especially for the wavelengths which are sometimes equal to the lower bounds. With the Traffic-Cutting method, the number of ADM's and wavelengths can not be greatly saved because of the limits of the star topology and the strictly nonblocking grooming. Only one more ADMs and no more wavelengths can be saved with this method when there are 15 nodes and 2 or 4 traffic patterns. While the other two splitting methods can get the same results and lead to 10 and 12 ADMs and 2 and 1 wavelengths savings, that amounts to 10.4% and 11.2% ADMs and 25% and 12.5% wavelengths savings respectively. So, they perform well when g is large enough. As was analyzed before, dividing a traffic flow will usually be an efficient way without the influence of the granularity, and by efficiently sending parts of a traffic flow through the spare capacity of the link, more ADMs and wavelengths can be saved.

In conclusion, although the search space may be larger with these three splitting methods, better grooming results can be achieved. When the number of nodes increases and traffic vary violently, the difference between splitting and non-splitting methods becomes more evident, which more effectively reveals the advantage of these splitting methods. So it proves that our algorithm can be applied to large scale star and tree networks. We also observed that, although it is hard to reach the lower bounds for arbitrary traffic patterns as mentioned in [5], the results got with splitting methods are closer or even equal to the lower bounds especially when both the numbers of nodes and traffic patterns are small. With the increase of either the number of traffic patterns or the number of nodes, it is harder and harder for the results to reach the lower bounds, but these three methods can usually save more ADMs and wavelengths than non-splitting method, and the last two methods performed better and more steadily in all the cases.

It can be found that among the three methods, the Traffic-Cutting method is usually not so efficient as the other two methods, especially in star networks. Though we can get good savings in almost all the cases with the Traffic-Dividing and the Synthesized-Splitting methods, it is obvious that the later leads to more savings at the cost of the delay of the traffic transmission and a more complex network virtual topology due to the multi-hopped connections. While the Traffic-Dividing method can avoid these problems, it is not as efficient in saving ADMs and wavelengths. Hence, trade-offs should be made between the complexity of network control and the savings of ADM's and wavelengths.

## 5. Conclusion

We have re-designed the three splitting methods proposed in [6] and applied them to the strictly nonblocking grooming in star and tree networks in this paper. Based on these three splitting methods, we proposed a novel genetic algorithm that can assign the same traffic demands at different traffic patterns to the same wavelengths by splitting appropriate traffic flows to achieve further savings. Computer simulation results showed that, in almost all the cases, these algorithms are more efficient in saving ADMs and wavelengths than the non-splitting ones. We analyzed the corresponding results and found that the Traffic-Dividing method is the best choice. As was shown in [6, 7], these three methods may perform better in the rearrangeably non-blocking grooming manner, and it will be our future work.